\documentclass[12pt,preprint]{aastex}

\usepackage{rotating}

\begin{document}

\title{A New Stellar Outburst Associated with the Magnetic Activities of the K-type Dwarf in a White-dwarf Binary}

\author{S.-B. Qian\altaffilmark{1,2,3,4}, Z.-T. Han\altaffilmark{1,2,3,4}, B. Zhang\altaffilmark{1,2,3,4}, M. Zejda\altaffilmark{5}, R. Michel\altaffilmark{6}, L.-Y. Zhu\altaffilmark{1,2,3,4}, E.-G. Zhao\altaffilmark{1,2,3,4}, W.-P. Liao\altaffilmark{1,2,3}, X.-M. Tian\altaffilmark{1,2,3,4} and Z.-H. Wang\altaffilmark{1,2,3,4}}

\altaffiltext{1}{Yunnan
Observatories, Chinese Academy of Sciences (CAS), P.O. Box 110, 650011
Kunming, P. R. China (e-mail: qsb@ynao.ac.cn)}

\altaffiltext{2}{Key laboratory of the structure and evolution of
celestial objects, Chinese Academy of Sciences, P.O. Box 110, 650011
Kunming, P. R. China}

\altaffiltext{3}{Center for Astronomical Mega-Science, Chinese Academy of Sciences, 20A Datun Road, Chaoyang Dis-
trict, Beijing, 100012, P. R. China}

\altaffiltext{4}{University of the Chinese Academy of
Sciences, Yuquan Road 19\#, Sijingshang Block, 100049 Beijing, P. R.
China}

\altaffiltext{5}{Department of Theoretical Physics and Astrophysics, Masaryk University, Kotl\'{a}\v{r}sk\'{a} 2, CZ-611 37 Brno, Czech Republic}

\altaffiltext{6}{Instituto de Astronom\'{i}a, Universidad Nacional Aut\'{o}noma de M\'{e}exico, Ensenada, Baja California, M\'{e}xico}

\begin{abstract}

1SWASP\,J162117.36$+$441254.2 was originally classified as an EW-type binary with a period of 0.20785\,days. However, it was detected to have undergone a stellar outburst on June 3, 2016. Although the system was latter classified as a cataclysmic variable (CV) and the event was attributed as a dwarf-nova outburst, the physical reason is still unknown. This binary has been monitored photometrically since April 19, 2016 and many light curves were obtained before, during and after the outburst. Those light and color curves observed before the outburst indicate that the system is a special CV. The white dwarf is not accreting material from the secondary and there are no accretion disks surrounding the white dwarf. By comparing the light curves obtained from April 19 to September 14, 2016, it was found that magnetic activity of the secondary is associated with the outburst. We show strong evidence that the $L_1$ region on the secondary was heavily spotted before and after the outburst and thus quench the mass transfer, while the outburst is produced by a sudden mass accretion of the white dwarf. These results suggest that J162117 is a good astrophysical laboratory to study stellar magnetic activity and its influences on CV mass transfer and mass accretion.

\end{abstract}


\section{Introduction}

1SWASP\,J162117.36$+$441254.2\,(=CSS\,J162117.3$+$441254=SDSS J162117.35$+$441254.1, hereafter J162117) was originally identified as an EW-type eclipsing binary by several authors (Palaversa et al. 2013; Lohr et al. 2013; Drake et al. 2014a). The derived orbital period (P=0.207852\,days) places the system near the short-period limit of contact binaries (Rucinski 1992, 2007; Qian et al. 2017). An outburst was reported by Drake et al. (2016) to have occurred on June 3, 2016 (UT = 10.8 h). The pre-discovery observations given by Maehara (2016) from May 28 to June 3, 2016 revealed that the object was already in outburst at $V$ = 13.13\,mag on June 1, 2016 (UT = 14.78 h). If the system is really a contact binary, the outburst could be explained as the beginning of a rare binary merger event similar to V1309 Sco (Tylenda et al. 2011; Zhu et al. 2016). However, the conclusion was ruled out by follow-up photometric monitoring and spectroscopic observations. Drake et al. (2016), using GALEX data as well, suspect that the outburst is more likely from an unusual cataclysmic variable (CV). Also, the spectroscopic emission lines observed by Scaringi et al. (2016) support that the outburst event is associated with accretion onto a compact object. Therefore J\,162117 is possibly a long-period CV above the period gap.
In addition to broad emission lines (Scaringi et al. 2016), the spectral observations obtained by Thorstensen (2016) show a strong contribution from a K-type secondary star. By analyzing the radial velocities, the masses of the compact object and donor star were estimated as $0.9$\,$M_{\odot}$ and $0.4$\,$M_{\odot}$, respectively.

After the report of the outburst from J\,162117 (Drake et al. 2016), the binary system was monitored continuously (Zejda \& Pejcha 2016; Pavlenko et al. 2016; Zola et al. 2016; Kjurkchieva et al. 2017). The phased light curve obtained by Zejda \& Pejcha (2016) during the outburst showed deep primary eclipses and shallower secondary eclipses. At the time, the depth of primary minima were decreasing, while the depth of secondary minima were increasing as the outburst faded. Finally, the system returned to its quiescence state during June 14-16, 2016 with its light curve back to being similar to EW-type variability (Zola et al. 2016). A 0.052-day variability with an amplitude of 0.1\,mag was reported by Pavlenko et al. (2016) that is superimposed on the out-of-eclipse light curve. These authors suspected that the variability could be related to the magnetic pole/poles of a white dwarf and that this could be a candidate  intermediate polar system. The outburst, with an amplitude of about 2 magnitudes in \textit{V} band and the EW-type light curve  in its quiescence state, make J\,162117 a very interesting target for further investigation. Although we know the outburst is associated with the accretion on a white dwarf and that the system may be an unusual CV, the physical reasons that produce this behaviour are still unknown.

Some CVs, e.g., the nova-like VY Scl-type variables and the strongly magnetic CVs (polars), usually show sudden dips in their brightness at irregular intervals of weeks to months. Low luminosity states have been explained as the coverage of dark spots near the $L_1$ point that then quench the mass transfer (Livio \& Pringle 1994; King \& Cannizzo 1998). Spots, produced at preferred longitudes, may well be due to tidal forces, and were theoretically predicted by Holzwarth \& Sch\"{u}ssler (2003). The Doppler image of the detached white-dwarf binary V471 Tau was conducted by Hussain et al. (2006) who found that the side of the star facing the white dwarf (around the $L_1$ point) was heavily spotted. These properties, that spots
are more preferred near the $L_1$ point, were also observed in a few CVs (e.g., Watson et al. 2006). However, we know relatively little about the influence of magnetic activity on the mass transfer and accretions in CVs. In the paper, we present photometric data of J162117 obtained by monitoring it from April 19 to September 14, 2016. We show that the mass accretion in J162117 is ceased by dark spots near the $L_1$ point before and after the outburst, while the outburst may be produced by an intermittent mass accretion on the white dwarf that is caused by the local magnetic activity of the secondary. These properties indicate that it may be a new type of optical outburst associated with the stellar activity.

\section{Photometric monitoring and the stellar outburst of J\,162117}

\begin{table}
\caption{Information of J\,162117, the comparison and check stars.}
\begin{tabular}{llll}
\hline\hline
              &      Variable               &  Comparison star             &    Check star          \\
\hline
 Name         &  CSS\_J162117.4+441254    &  LAMOST241916119        &   LAMOST241916114      \\
RA(2000)      &  245.322395               &  245.352570             &   245.433867           \\
Dec(2000)     &  +44.215008               &  +44.220438             &   +44.181152           \\
U             &  $15.980(\pm0.010)$            &  $16.815(\pm0.014)$          &   $15.710(\pm0.013)$        \\
B         &  $16.000(\pm0.007)$            &  $16.116(\pm0.006)$          &   $15.730(\pm0.016)$        \\
V         &  $15.217(\pm0.005)$            &  $15.182(\pm0.004)$          &   $15.142(\pm0.008)$        \\
R         &  $14.606(\pm0.005)$            &  $14.658(\pm0.006)$          &   $14.805(\pm0.008)$        \\
I         &  $14.050(\pm0.005)$            &  $14.188(\pm0.005)$          &   $14.411(\pm0.005)$        \\
J         &  $13.462(\pm0.027)$            &  $13.513(\pm0.026)$          &   $13.975(\pm0.024)$        \\
H         &  $12.875(\pm0.023)$            &  $13.095(\pm0.025)$          &   $13.540(\pm0.029)$        \\
K         &  $12.763(\pm0.025)$            &  $12.943(\pm0.025)$          &   $13.504(\pm0.036)$        \\
\hline\hline
\end{tabular}
\end{table}

\begin{figure}
\begin{center}\label{fig1}
\includegraphics[angle=0,width=0.75\columnwidth]{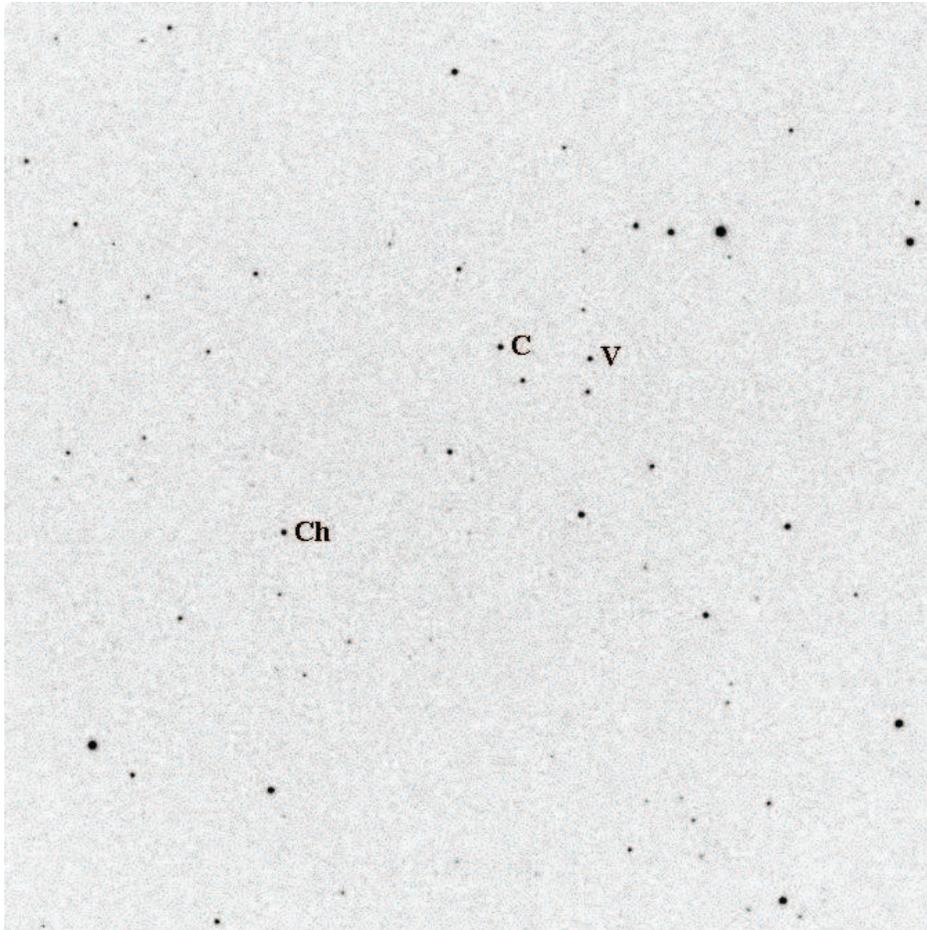}
\caption{A finding chart of the variable star (J\,162117). "C" refers to the comparison star, while "Ch" to the check star.}
\end{center}
\end{figure}

\begin{figure}
\begin{center}\label{fig2}
\includegraphics[angle=0,width=0.99\columnwidth]{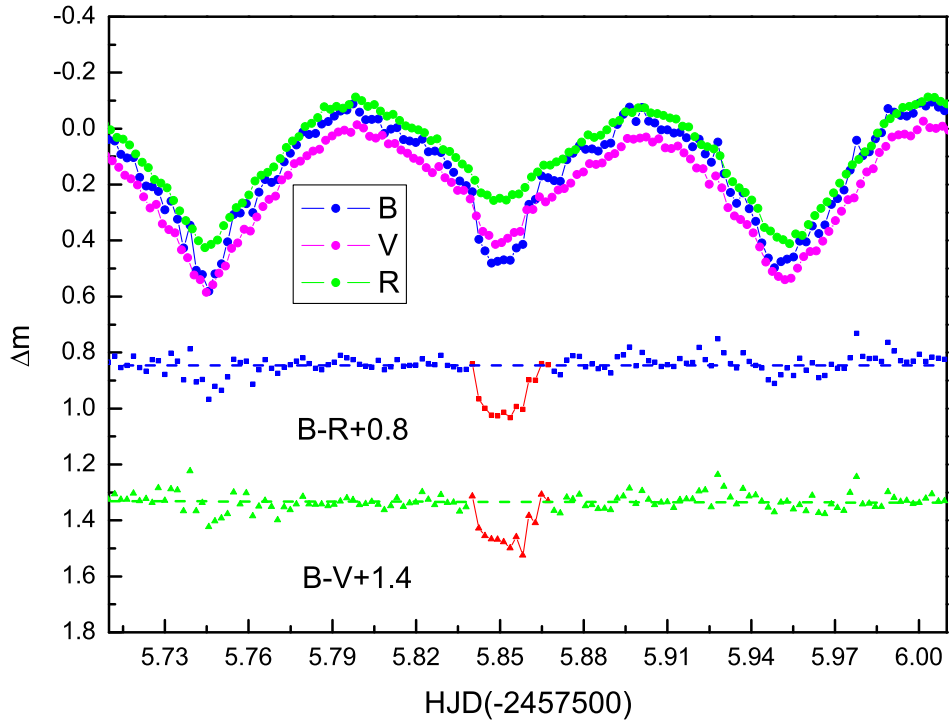}
\caption{Photometric light curves in \textit{BVR} bands observed on April 27, 2016. Also shown in the figure as squares and triangles are the $B-R$ and $B-V$ colors, respectively. Two shoulders in $B$ and $V$ band light curves are clearly seen at both sides of the brightness minimum around HJD\,2457505.85 indicating that there are no accretion disks surrounding the white dwarf. This is supported by the $B-R$ and $B-V$ color curves (see text for details). The \textit{BVR} light curves are asymmetric and some small optical flares in B-band light curves are visible.}
\end{center}
\end{figure}

\begin{figure}
\begin{center}\label{fig3}
\includegraphics[angle=0,width=0.99\columnwidth]{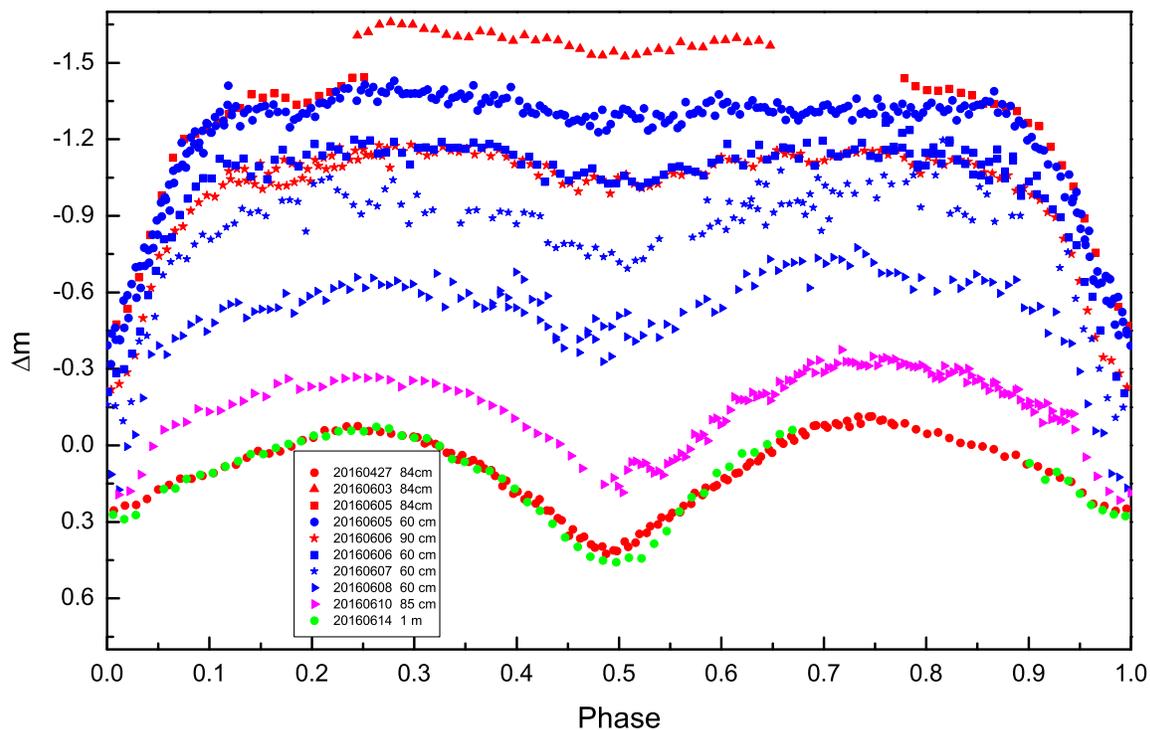}
\caption{Comparison of the light curves in $R$ band obtained during the outburst and outside the outburst from April 27 to June 14, 2016. The light curves during the outburst are varying rapidly, while those obtained before and after the outburst are nearly overlapping and are showing asymmetric (O'Connell effect).}
\end{center}
\end{figure}

\begin{table}\label{tab2}
\caption{The log of photometric monitor for J\,162117 in 2016. Start time is HJD-2\,457\,500.}
\centering
\footnotesize
\begin{tabular}{lllll}
\hline\hline
Date           &Filters             & Start time  & Duration (Hours) &  Telescopes\\\hline
Mar 19    &\textit{BVR}                 &-33.0620        &2.40       & 84 cm \\
Mar 21    &\textit{BVR}                 &-31.0632        &1.37       & 84 cm \\
Apr 27    &\textit{BVR}                 &5.7052          &7.35       & 84 cm \\
Jun 03    &\textit{BVR}                 &42.8993         &2.05       & 84 cm \\
Jun 05    &\textit{BVR}                 &44.8811         &2.35       & 84 cm \\
Jun 05    &$R$                   &45.3309         &5.88       & 60 cm \\
Jun 06    &\textit{BVR}                 &46.4068         &5.80       & 90 cm \\
Jun 06    &\textit{RI}                  &46.3275         &5.84       & 60 cm \\
Jun 07    &\textit{RI}                  &47.3293         &5.73       & 60 cm \\
Jun 08    &\textit{RI}                  &48.3292         &5.68       & 60 cm \\
Jun 10    &\textit{BVR}                 &50.0813         &5.71       & 85 cm \\
Jun 12    &\textit{RI}                  &52.3853         &3.67       & 60 cm \\
Jun 13    &\textit{RI}                  &53.3548         &5.07       & 60 cm \\
Jun 14    &\textit{RI}                  &54.3376         &3.83       & 60 cm \\
Jun 14    &\textit{BVR}                 &54.0518         &3.83       & 1 m \\
Jun 15    &\textit{VR}                  &55.1051         &5.32       & 85 cm \\
Jun 15    &\textit{RI}                  &55.3347         &4.47       & 60 cm \\
Jun 17    &\textit{RI}                  &56.5216         &0.64       & 60 cm \\
Jun 18    &$R$                   &58.3403         &5.69       & 60 cm \\
Jun 22    &$R$                   &62.3363         &5.76       & 60 cm \\
Jun 23    &\textit{RI}                  &63.3405         &5.44       & 60 cm \\
Jun 23    &\textit{BVR}                 &63.1080         &3.10       & 1 m \\
Jun 24    &\textit{BVR}                 &64.1550         &3.02       & 1 m \\
Jun 24    &\textit{RI}                  &64.3538         &5.18       & 60 cm \\
Jun 25    &$R$                   &65.3344         &1.78       & 60 cm \\
Jun 26    &$R$                   &66.3477         &0.42       & 60 cm \\
Jun 27    &\textit{VR}                  &67.4406         &2.31       & 60 cm \\
Jun 28    &$R$                   &68.3431         &5.42       & 60 cm \\
Jun 29    &\textit{VR}                  &69.3880         &1.89       & 60 cm \\
Jun 29    &$R$                   &69.3110         &1.23       & 1 m \\
Jul 01    &\textit{VR}                  &71.3350         &5.78       & 60 cm \\
Jul 04    &\textit{VR}                  &74.3348         &4.01       & 60 cm \\
Jul 05    &\textit{VR}                  &75.3315         &5.97       & 60 cm \\
Jul 08    &\textit{BVR}                 &78.0390         &1.59       & 1 m \\
Jul 09    &\textit{BVR}                 &79.0334         &4.35       & 1 m \\
Jul 24    &\textit{BVR}                 &94.0791         &1.72       & 1 m \\
Aug 19    &\textit{BVR}                 &120.0268        &2.43       & 70 cm \\
Aug 23    &\textit{BVR}                 &124.0096        &3.3        & 1 m \\
Sep 13    &\textit{VR}                  &144.9973        &0.68       & 1 m \\
Sep 14    &\textit{VR}                  &146.0052        &2.33       & 1 m \\
\hline
\end{tabular}
\end{table}

\begin{figure}
\begin{center}\label{fig4}
\includegraphics[angle=0,width=0.9\columnwidth]{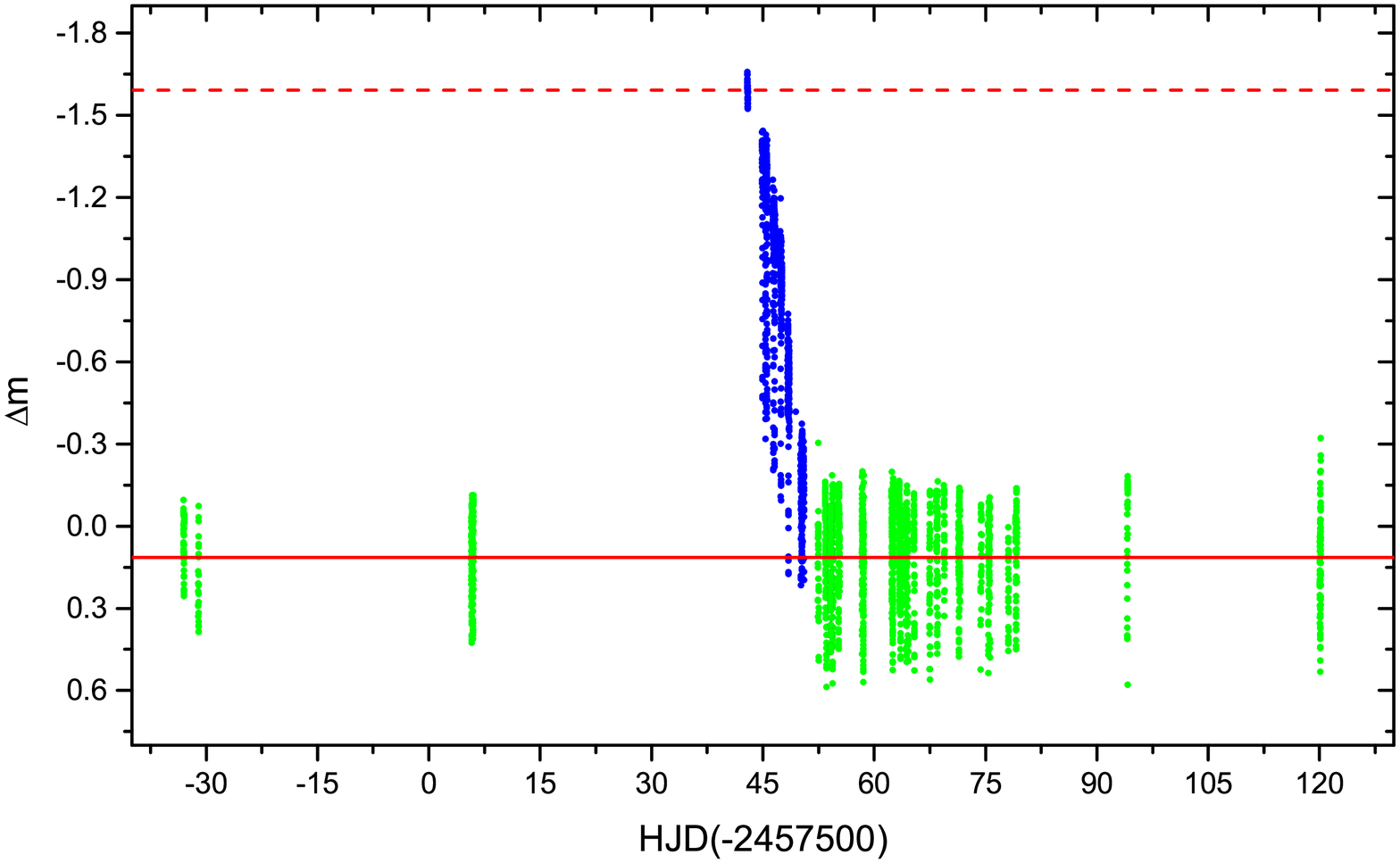}
\caption{The brightness variation of J162117 in $R$ band observed from March 19 to September 14, 2016. Blue dots represent the brightness change during the outburst, while green ones refer to the brightness outside the outburst.}
\end{center}
\end{figure}

Due to its unusually short-period and showing EW-type light variation (Palaversa et al. 2013; Lohr et al. 2013; Drake et al. 2014a), J\,162117 was included in our observational list of contact binary stars below or near the short-period limit (Qian et al. 2014a; Qian et al. 2015a,b; Jiang et al. 2015). We started to observe the binary on March 19, 2016, by using the 84-cm telescope in Mexico.
The observations were carried out with the 0.84-m f/15
Ritchey-Chr\'etien telescope at OAN-SPM Baja California, the
Mexman filter-wheel and the $Spectral Instruments$ CCD detector (a
deep depletion e2v CCD42-40 chip which has a 2048$\times$2048 13.5
$\mu$m square pixel array, a gain of 1.32 e$^-$/ADU, and a readout
noise of 3.4 e$^-$). 2$\times$2 binning was used during all these
observations with exposure times set at 60 sec for the filter $B$
integrations, 30 sec for $V$ and 20 sec for $R$. Flat field and bias
frames were also acquired during all the observing runs.
Original CCD images were reduced by using PHOT (measure magnitudes for a list of stars) of the aperture photometry package of IRAF.

As shown in Fig. 1, two stars near J\,162117 were chosen as the comparison and the check stars, respectively. Their coordinates and the calibrated magnitudes together with the corresponding errors are listed in Table 1. On March 19 and 21, 2016, only data around eclipses were obtained, while complete light curves in \textit{BVR}-bands were observed and are shown in Fig. 2. Also displayed in the figure are the $B-V$ and $B-R$ color curves. These pre-outburst light and color curves are very useful for understanding the properties of J\,162117. The data displayed in Fig. 2 are available at the website\footnote{http://search.vbscn.com/J162117.Fig2data.txt} via the internet.

Although the light curves resemble those of EW types, the details are quite different. The ingress and egress of the shallower eclipsing minima are visible in the $B$ and $V$-band light curves. The two shoulders around the minimum are more clearly seen in the $B$-band light curve and its bottom is nearly flat. These properties indicate that this minimum is caused by the eclipse of a compact object by a normal cool star, revealing that it is a WD+MS binary system. This is consistent with the conclusion derived by previous investigators (Drake et al. 2016). Meanwhile, as shown in Fig. 2, the eclipse minimum in $B$ band is deeper than those in $V$ and $R$ bands, indicating that the WD is hotter than its main-sequence companion. Therefore, this shallower minimum should be the primary one that corresponds with the eclipse of the primary component, i.e., a white dwarf. During the outburst, it becomes deeper, while the other one nearly disappears.

Because of the peculiar observational properties of J\,162117, after the outburst was reported on June 3, 2016 by Drake et al. (2016), we continue to monitor the target by using the 84-cm and 90-cm telescopes in Mexico. The comparison of the light curves obtained from April 27 to June 14 is shown in Fig. 3. The data obtained on June 3 indicate that the secondary minimum became shallower and nearly disappeared. To get more photometric data, J\,162117 was then monitored continuously by using several small telescopes in the Czech Republic and in China. The log of the photometric monitoring for the system is shown in Table 2 where the dates and the filters used are listed in the first and the second columns. Those listed in the third and the fourth columns are the start time (in HJD-2\,457\,500) and the duration of the observations.
As shown in Fig. 3, during the outburst the primary eclipses are very deep, while the secondary minima are shallow. As reported by Zejda \& Pejcha (2016), the depth of primary minimum decreases, while the depth of secondary minimum increases as the outburst fades. The system returned to a quiescence state on June 14, 2016, and the light curves obtained before and after the outburst are nearly overlapping.

The total light curves in $R$ band observed from March 19 to September 14, 2016, are displayed in Fig. 4 where blue dots refer to the light curves obtained during the outburst while green ones refer to those observed outside the outburst. As shown in Fig. 4, the amplitude of the outburst in $R$ band is larger than 1.71\,mag. It takes about 11 days that the system returns to the quiescence brightness state. Before and after the outburst, the brightness levels are nearly the same. All of the R-band photometric data displayed in Fig. 4 are available at the website\footnote{http://search.vbscn.com/J162117.Fig4data.txt} via the internet.

\section{Discussions and conclusions}

During the outburst of J\,162117, the spectral observations showed some properties of CVs, i.e., the broad, two-peaked emission lines $H_{\alpha}$ (FWHM corresponding to 1500 km/s), $H_{\beta}$, and HeII\,4686. These are indications of the accretion on a white dwarf (Scaringi et al. 2016). Moreover, the strong rotational disturbance of the emission lines in the eclipse indicates the presence of a rapidly-rotating disk (Thorstensen 2016). Therefore, after the merging of a contact binary was ruled out to explain the outburst of J\,162117, this event was attributed to a dwarf-nova outburst (Scaringi et al. 2016; Kjurkchieva et al. 2017). However, when comparing with known CVs, J\,162117 shows several peculiarities (Kjurkchieva et al. 2017), namely a deeper eclipse at outburst than at quiescence (by a factor 2.8) and outburst amplitudes at the lowest limit of dwarf nova eruptions. These properties reveal that J\,162117 is not a normal CV.

The binary system was monitored photometrically from April 19 to September 14, 2016, by using several small telescopes in the world. We were lucky to obtain several high-precision light curves before the outburst. As shown in Fig. 2, there are two shoulders around the primary minima in the B- and V-band light curves. The two minima show a clearly U-type shape and their bottoms are nearly flat. The color curves plotted in the figure also show a U-type shape around the primary minima, while the color curves are flat outside the eclipses. All of these properties suggest that there are no accreting disks around the white dwarf and thus the white dwarf is not accreting material from the cool secondary at the pre-outburst quiescent state.

Apart from the eclipse of the white-dwarf component, the \textit{BVR} light curves shown in Fig. 2 are dominated by ellipsoidal variability in a binary star as well as the magnetic activity of the K-type secondary. As displayed in Fig. 2, some small flare-like events are visible in B-band light curves, while these events are not seen in $V$ and $R$ band light curves. These are general properties of optical flares (e.g., Qian et al. 2012). Moreover, as shown in Table 1, the errors of the B-band magnitudes for J162117 and the comparison star are about $\pm0.006$ indicating those B-band flares are true.
The light curves in the figure are asymmetric and showing O'Connell effect. Meanwhile, the depth of the secondary minimum (at phase 0.5) is deeper than the primary one (at phase 0.0). These observed properties could be explained by the secondary being covered by dark spots near the inner Lagrange point $L_1$ where the mass is expected to flow onto the primary. The fact that regions near the $L_1$ point are heavily spotted has been observed in the pre-CV V471 Tau and a few CVs (e.g. Hussain et al. 2006; Watson et al. 2006). This type of dark spots, that may be caused by tidal forces and produced at preferred longitudes, was theoretically predicted by Holzwarth \& Sch\"{u}ssler (2003). The coverage of dark spots around the $L_1$ region could cease the mass transfer and thus quench the mass accretion of the white dwarf (e.g. Livio \& Pringle 1994; Qian et al. 2014b, 2015c,d). This is the reason why there are no accreting disks around the white dwarf before the outburst.

\begin{figure}
\begin{center}
\includegraphics[angle=0,width=0.9\columnwidth]{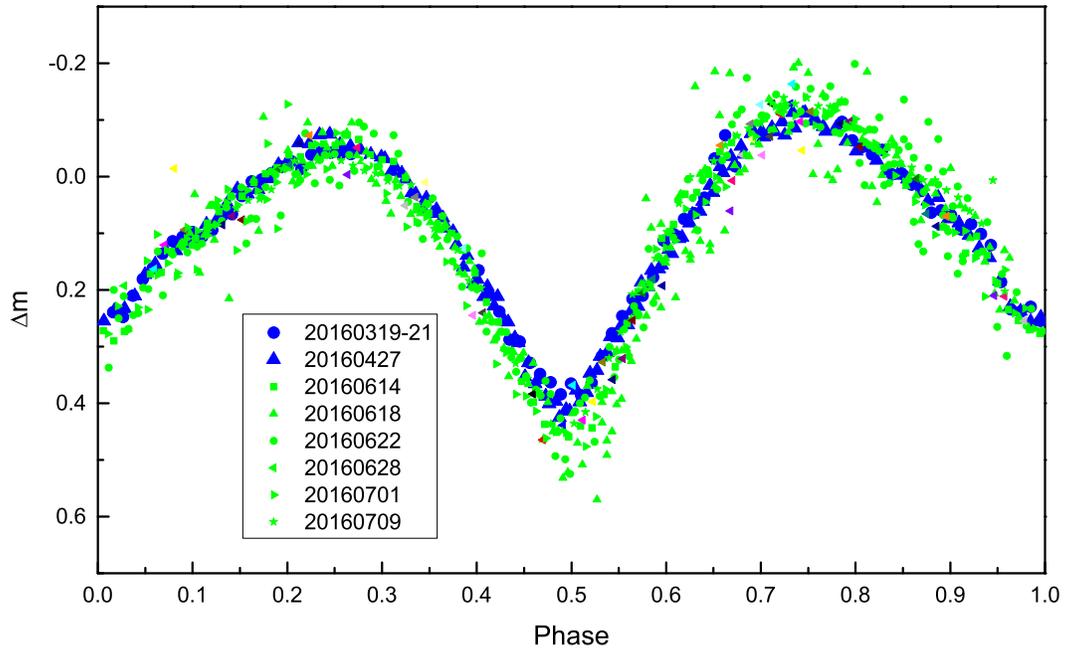}
\caption{Comparison of several light curves in R band obtained from March 18 to July 9, 2016. All of the light curves were observed outside the stellar outburst. Those light curves are stable and are showing asymmetry (O'Connell effect).}
\end{center}
\end{figure}

\begin{figure}
\begin{center}
\includegraphics[angle=0,width=0.9\columnwidth]{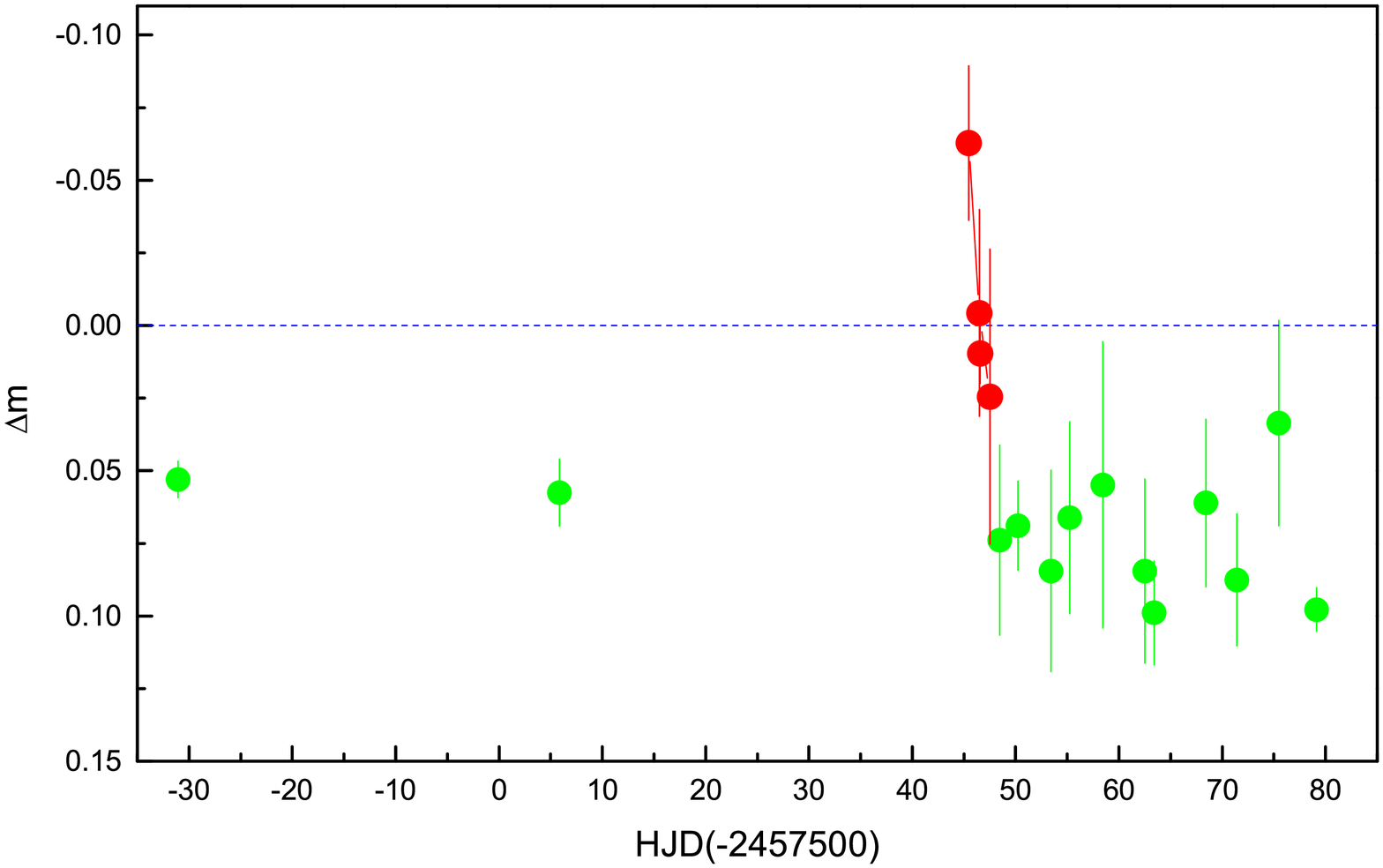}
\caption{Variation of the magnitude difference between the two maxima (the O'Connell effect). Red solid dots refer to the data obtained during the outburst, while green dots
to those observed outside the outburst. It is shown that the O'Connell effect varies rapidly during outburst, while it is stable outside the outburst within the error.}
\end{center}
\end{figure}

The comparison of some $R$-band light curves, observed outside the stellar outburst from March 18 to July 9, 2016, is shown in Fig. 5. The light curves are nearly overlapping and no changes in their shapes are noticeable. All of them are asymmetric (the O'Connell effect) and are showing deeper secondary minima.
Meanwhile, the variation of the O'Connell effect (the magnitude difference between the two maxima) is displayed in Fig. 6 where red solid dots refer to the data observed during the outburst, while green ones refer to those obtained outside the outburst. The magnitudes at the two maxima were determined by averaging the data around phases 0.25 and 0.75, respectively. As shown in the figure, the O'Connell effect is varying rapidly during the outburst, while it is stable outside the outburst within the error. All of the observations of J162117 reveal that there are dark spots near the $L_1$ point at those quiescent states. The dark spots on the K-type secondary cease the mass accretion on the white dwarf as well as cause those deeper secondary minima and the O'Connell effect. During the outburst, the dark spots at the $L_1$ point may disappear and the sudden mass accretion on the white dwarf cause the stellar outburst. The observed broad, two-peaked emission lines (e.g., $H_{\alpha}$) and the strong rotational disturbance of the emission lines during the outburst could be explained by the accretion of the white-dwarf component.

Here we speculate that the K-type secondary is only marginally filling the critical Roche lobe. The coverage of dark spots on the secondary could cause it to expand (e.g, Chabrier et al. 2007) and cause materials of the cool secondary to overflow from the Roche lobe. However, J162117 was monitored for many years by the SuperWASP project and had a stable light curve showing the O'Connell effect, as observed in the pre-outburst light curves in Fig. 2. This implies magnetic spot coverage for a long time prior to the outburst indicating
that the expanding may be a long-term property of the cool secondary. After it expands to fill the Roche lobe completely, the white dwarf is then accreting mass from the secondary suddenly and producing the outburst. During the outburst, the dark spots around the $L_1$ region disappear because of the overflow of material onto the hot white dwarf. Finally, the shrinking of the secondary due to the disappearance of the dark spots causes the accretion rate to decrease and the system returns to its quiescent state. Meanwhile, dark spots near the $L_1$ region are produced again due to tidal forces, as predicted by Holzwarth \& Sch\"{u}ssler (2003) .

The present investigation indicates that J162117 is not a normal CV and the stellar eruption detected on June 3, 2016 is not a dwarf-nova outburst, because no lasting accretion disks are around the white dwarf. It is shown that the magnetic activity of the secondary may be associating with the outburst. During the quiescent states, the local dark spots near the $L_1$ point cease the accretion onto the white dwarf, while the optical outburst is produced by an intermittent mass accretion. The mass accretion during the outbursts may be caused by the expanding of the secondary via the presence of the dark spots. In this way, both the greater eclipse depth at outburst than at quiescence and the low outburst amplitudes (only about 2 mag)
could be explained. The eruption event in J\,162117 may be a new optical outburst associated with the stellar activity. These results make J\,162117 a very interesting target for future investigations on stellar magnetic activity and its influence on CV mass transfer and mass accretion.

\acknowledgments{The work is partly supported by the Chinese Natural Science Foundation (Nos. 11325315, 11573063, and 11133007). MZ was supported by the project GA \v{C}R 16-01116S. New observations were obtained with the 1.0-m and the 70-cm telescopes in YNOs and the 85-cm telescope at the Xinglong station of NAOs in China, the 84-cm and 90-cm telescopes in Mexico, and the 60-cm Newtonian telescope of Masaryk University in the Czech Republic.}

\end{document}